\documentclass[aps,prd,nofootinbib,floats,
floatfix,preprintnumbers,groupedaddress]{revtex4}
\usepackage{comment}
\usepackage{graphicx}
\usepackage{amsmath, mathrsfs}
\usepackage{float}
\usepackage{subfigure}
\usepackage[usenames]{color}
\usepackage{amssymb}
\usepackage{bbm}
\usepackage{csquotes}
\usepackage{color}
\usepackage{psfrag}
\usepackage{graphicx}
\usepackage{orcidlink}
\usepackage{caption,subcaption}
\usepackage{marginnote}
\usepackage[utf8]{inputenc}
\usepackage{tikz}
\usetikzlibrary{positioning,decorations.pathmorphing}

\usepackage{epstopdf}
\newcommand{\bea}{\begin{aligned}}
\newcommand{\eea}{\end{aligned}}
\newcommand{\beq}{\begin{equation}}
\newcommand{\eeq}{\end{equation}}

\newcommand{\bse}{\begin{subequations}}
\newcommand{\ese}{\end{subequations}}

\usepackage{hyperref}
\hypersetup{
     colorlinks   = true,
     citecolor    = red,
     linkcolor    = blue,
     urlcolor     = blue,
}

\newcommand{\bmm}{\begin{multline}}
\newcommand{\emm}{\end{multline}}

\begin{document}
\title{\Large{Noncommutative Geometry and the Thermodynamic Fate of Black Holes\\}}
\author{\bf Ankit Anand$^1$\orcidlink{0000-0002-8832-3212}, Anshul Mishra$^2$, and Aditya Singh$^3$\orcidlink{0000-0002-2719-5608} \\}
\email{Anand@iitk.ac.in, anshulmishra2025@gmail.com, 24pr0148@iitism.ac.in}
\affiliation{$^1$Department of Physics, Indian Institute of Technology Kanpur, Kanpur 208016, India.\\
$^2$Department of Physics, Indian Institute of Technology Madras, Chennai 600036, India.\\
$^3$Department of Physics, Indian Institute of Technology (Indian School of Mines), Dhanbad, Jharkhand-826004, India.
}



\pagenumbering{arabic}

\renewcommand{\thesection}{\arabic{section}}
\begin{abstract}
\begin{center}
    \hspace{1cm}
\end{center}
We study the thermodynamics of black holes in the framework of non-commutative geometry, where spacetime fuzziness is modelled by smeared Lorentzian distributions. Corrected black hole solutions with this quantum fuzziness are obtained, and their thermodynamic analysis is performed. We show that the conventional first law of black hole thermodynamics is violated since the entropy deviates from the Bekenstein-Hawking form. Introducing a correction to the mass restores consistency, yielding a modified first law compatible with Bekenstein-Hawking entropy. Next, we investigate the effects of spacetime non-commutativity on the thermodynamic universality of these black holes. We demonstrate that non-commutativity modifies the standard universality relations of black holes and can induce thermodynamic stability by altering the underlying microscopic interactions. Our results suggest that quantum features of spacetime can have significant macroscopic consequences for black hole thermodynamics.

\end{abstract}


\maketitle


\section{Introduction}
Geometric frameworks have significantly enriched the understanding of black hole thermodynamics by providing deeper insights into their underlying statistical behavior. Notably, the scalar curvature of the thermodynamic parameter space acts as a sensitive indicator of phase transitions and serves as a powerful diagnostic of the microscopic interactions governing black hole microstructure. Black holes provide a natural arena for exploring the interplay between quantum mechanics and general relativity, particularly in regimes where classical spacetime descriptions break down. Their thermodynamic properties—such as entropy, Hawking radiation, and microstructure—are deeply connected to the quantum nature of gravity. In recent years, considerable attention has been devoted to understanding quantum black holes and their dynamics within various theoretical frameworks \cite{fursaev1995temperature, xiao2022logarithmic, battista2024quantum, calmet2021quantum, wang2025dynamical, Munch_2023, Mele_2022, Bodendorfer_2019, Koch_2016, PhysRevD.97.024027, rincon2018scale, panotopoulos2021quasinormal}. Among these, non-commutative geometry has emerged as a promising approach, wherein spacetime coordinates are promoted to noncommuting operators, introducing a fundamental minimal length scale \cite{doplicher1995quantum}. This non-commutativity leads to a fuzziness of spacetime, effectively replacing point-like structures with smeared distributions, such as Gaussian-sourced matter profiles. The resulting "spacetime fuzziness" has profound implications for black hole physics: it smooths out curvature singularities, modifies horizon structure, and leads to finite thermodynamic quantities even at small scales \cite{nicolini2006noncommutative, Smailagic:2003yb}. These geometric deformations also influence the thermodynamics universality relation, make the black holes thermodynamically stable by changing its microscopic interactions captured by thermodynamic scalar curvature and its curvature through the lens of Ruppeiner geometry, which provides insight into the nature of microscopic interactions within black holes. In this context, the thermodynamic scalar curvature—interpreted as a measure of statistical interactions—captures the imprints of non-commutativity, enabling a deeper understanding of the quantum-corrected microstructure of spacetime.

Non-commutative geometry offers a mathematically robust approach to incorporating quantum effects into the fabric of spacetime by generalizing the notion of a smooth manifold. In this framework, spacetime coordinates are promoted to noncommuting operators obeying the algebra
\begin{equation}\label{NCRelation}
[y^\mu, y^\nu] = i\Theta^{\mu\nu} \ ,
\end{equation}
where $\Theta^{\mu\nu}$ is a real, constant antisymmetric tensor that characterizes the non-commutative deformation of spacetime \cite{doplicher1995quantum, seiberg1999string, smailagic2003feynman,Lizzi:1997yr}. This commutation relation introduces a natural ultraviolet cutoff by enforcing a fundamental length scale, effectively leading to a "fuzzy" spacetime wherein the classical concept of sharply localized points is replaced by non-local structures. The implications of non-commutative geometry for gravitational physics, particularly in the context of black hole solutions, have been widely investigated \cite{mann2011cosmological, Modesto:2010rv, nicolini2006noncommutative, lopez2006towards, calmet2006second, aschieri2005gravity,Calmet:2005qm}. A central consequence is the regularization of the energy-momentum source. Instead of modeling black holes with a point-like singular mass, non-commutative-inspired models replace the delta-function source with a smeared mass distribution. Two frequently employed profiles are the Lorentzian and Gaussian distributions, and they are 
\begin{eqnarray}\label{RhoL}
\rho_{_L}(r) = \frac{M \sqrt{\Theta}}{\pi^{3/2}(r^2 + \pi\Theta)^2} \;\;\;\;\;\;\;\;\;\;\;;\;\;\;\;\;\;\;\;\;\; 
\rho_{_G}(r) = \frac{M}{(4\pi \Theta)^{3/2}} e^{-\frac{r^2}{4\Theta}} \ ,
\end{eqnarray}
where $M$ is the total mass and $\Theta$ denotes the non-commutative parameter. These smeared sources reflect the spacetime fuzziness and lead to black hole metrics that are free from curvature singularities at the origin. Lorentzian-smeared black holes also display qualitatively similar features and have been studied for their thermodynamics, phase structure, and quasinormal mode spectra \cite{wang2024thermodynamics, campos2022quasinormal, araujo2024effects}. These models demonstrate how non-commutative geometry can serve as a physically motivated tool to probe the quantum gravitational regime, effectively smoothing out ultraviolet divergences and offering new insights into black hole microstructure.

The introduction of geometric concepts into thermodynamics began with~\cite{Weinhold:1975xej}, who proposed a Riemannian structure on the thermodynamic phase space by defining a metric as the Hessian of the internal energy with respect to extensive variables. Shortly thereafter~\cite{Ruppeiner:1995zz, Ruppeiner:1979bcp}, motivated by the foundational principles of statistical mechanics and the Boltzmann entropy formula, formulated an alternative geometry in which entropy plays the role of the thermodynamic potential. In this framework, the thermodynamic state space acquires a Riemannian structure whose scalar curvature, denoted by $\mathcal{R}$, encodes key information about fluctuations and phase transitions. This thermodynamic curvature has been found to exhibit critical behavior near phase transitions, diverging at critical points and thereby providing a geometric signature of macroscopic instabilities. Ruppeiner geometry has been extensively applied to a wide class of systems, including ideal classical gases, van der Waals fluids, low-dimensional spin systems, and quantum gases \cite{Ruppeiner:1981znl, Janyszek:1989zz, Janyszek:1990wdh}, thereby revealing universal features across diverse models. A notable interpretation of the scalar curvature is that it reflects the statistical interaction among microscopic constituents: negative curvature is associated with attractive interactions (as in Bose gases), positive curvature with repulsive interactions (as in Fermi gases), and zero curvature with non-interacting systems such as the classical ideal gas. Physically, the line element in Ruppeiner geometry measures the thermodynamic "distance" between nearby fluctuation states, with longer distances implying lower probabilities of spontaneous transitions. This interpretation bridges thermodynamic geometry and statistical mechanics by connecting fluctuation theory with curvature. Consequently, the sign and magnitude of $R$ provide valuable insight into the nature and strength of micro-interactions, a feature that has been employed to characterize the microscopic structure of complex thermodynamic systems \cite{Oshima:1999mjq, RuppeinerGeorge}. Furthermore, it has been conjectured that the Ruppeiner scalar curvature is related to the correlation length $\xi$ of the system through a scaling relation of the form $ R \sim \kappa\, \xi^{\bar{d}},$ where $\kappa$ is a dimensionless constant and $\bar{d}$ denotes the effective spatial dimensionality. Near criticality, where $\xi \to \infty$, the scalar curvature diverges accordingly, marking the onset of a phase transition. Thus, Ruppeiner geometry provides a macroscopic diagnostic for probing microscopic interactions, serving as a thermodynamic analogue of the inverse problem in statistical mechanics. Rather than deriving thermodynamic properties from a microscopic model, this geometric approach allows one to infer aspects of the microstructure directly from macroscopic thermodynamic data—offering a powerful and general framework for analyzing critical phenomena and phase behavior.



\section{Black hole in non-commutative geometry}
In the presence of a cosmological constant, Einstein’s field equations take the form
\begin{equation}\label{Einstein equation}
R_{\mu\nu} - \frac{1}{2} g_{\mu\nu} R + \Lambda g_{\mu\nu} = 8\pi T_{\mu\nu}, 
\end{equation}
where $R_{\mu\nu}$ is the Ricci tensor, $R$ the Ricci scalar, $g_{\mu\nu}$ the spacetime metric, and $T_{\mu\nu}$ the energy–momentum tensor of matter. In a static, spherically symmetric setting, the line element may be written as
\begin{equation}\label{Ansatz}
\mathrm{d}s^2 = -f(r) \mathrm{d}t^2 + \frac{\mathrm{d}r^2}{f(r)} + r^2 \mathrm{d}\Omega^2 \ , 
\end{equation}
with the metric function $f(r)$ to be determined from the field equations. Inserting this ansatz into Eq. \eqref{Einstein equation}, the $tt$-component of Einstein equations takes the form 
\begin{eqnarray}
    \frac{1}{2} f(r) \left(f''(r)+\frac{2 f'(r)}{r}\right) -\frac{1}{2} f(r) \left(\frac{r^2 f''(r)+4 r f'(r)+2 f(r)-2}{r^2} + 2\Lambda\right) = 8\,\pi\, T_{tt} \ .
\end{eqnarray}
Rearranging this reduces to a single radial equation for $f(r)$, given by
\begin{equation}\label{f(r) Integral}
f(r) = 1 + \frac{1}{r} \int_0^r \left[8\pi y^2 T^t_t(x) - \Lambda x^2 \right] \mathrm{d}x \ . 
\end{equation}
Within the framework of non-commutative geometry, spacetime coordinates are promoted to operators obeying a non-trivial commutation relation~\eqref{NCRelation}. This modification is often implemented by replacing the classical Dirac delta mass density with a smeared, regular distribution. Now, using the distributions~\eqref{RhoL}, we will first compute the metric function for the Lorentzian smearing function. Finally, putting Eq.~\eqref{RhoL} into Eq.~\eqref{f(r) Integral} and performing the integration yields a non-commutative-corrected metric function of the form
\begin{eqnarray}\label{metric_function}
    f(r) = 1+ \frac{32 \alpha  M}{\pi ^2 \alpha ^2+64 r^2}-\frac{4 M}{\pi  r} \tan ^{-1}\left(\frac{8 r}{\pi  \sqrt{\alpha ^2}}\right)-\frac{\Lambda  r^2}{3} \ .
\end{eqnarray}
The metric recovers the Schwarzschild-AdS solution in the limit $\theta \to 0$ or $\alpha \rightarrow 0$\footnote{By expanding one can easily verify \begin{equation*}
f(r) = 1-\frac{2 M}{r}-\frac{\Lambda  r^2}{3}+\frac{\alpha  M}{r^2}-\frac{\pi ^2 \alpha ^3 M}{96 r^4}+ \mathcal{O}(\alpha^4)  \ .
\end{equation*}} 
where, $\alpha =8 \sqrt{\Theta/\pi}$. The terms involving $\alpha$ and $\alpha^3$ represent leading-order quantum gravitational corrections from spacetime fuzziness. These conditions define the admissible parameter space for the existence of non-commutative Schwarzschild-AdS black holes. The non-commutative corrections encoded in $\alpha$ significantly alter the near-horizon geometry, affect the thermodynamic behavior, and regularize the small-scale structure of the black hole spacetime.

\subsection*{Analysis of the horizons}

\begin{figure}[h]
    \centering
    \includegraphics[width=0.45\textwidth, height=6cm]{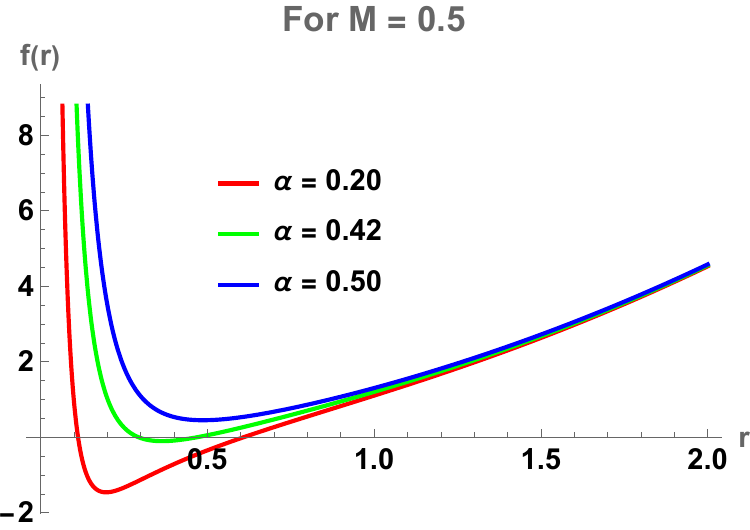}
    \includegraphics[width=0.45\textwidth, height=6cm]{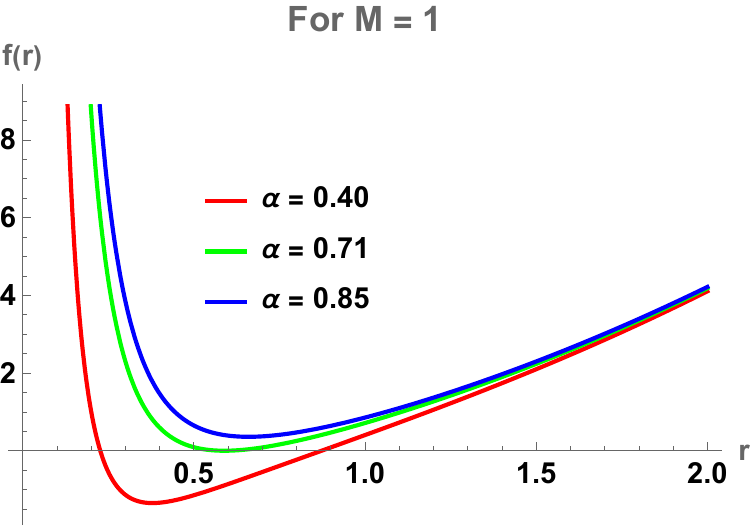}
    \vspace{5pt}
    \caption{The lapse function $f(r)$ for $M=1$ and different values of $\alpha$. The qualitative behavior of the horizons is the same as in the $M=0.5$ case, but with a larger outer horizon radius due to the larger mass.}
    \label{Fig:Metric Plot}
\end{figure}

The event horizon(s) are determined by the zeroes of the lapse function $f(r)$ as in Eq.~\eqref{metric_function} but perturbatively\footnote{The non-commutative parameter $\alpha$ characterizes the scale at which spacetime non-commutativity becomes relevant, typically assumed to be much smaller than the black hole horizon radius ($\sqrt{\alpha} \ll r_h$). Therefore, its effects can be treated as small corrections to the classical solution. Expanding perturbatively in $\alpha$ allows us to capture the leading-order non-commutative corrections analytically while keeping the calculations tractable.} up to $\alpha$ only. Taking upto first order of $\alpha$ we have to solve the quartic equation
\begin{equation}
F(r,\alpha) \;=\; r^{4}+\ell^{2}r^{2}-2\ell^{2}M r + \ell^{2} M \alpha = 0 \ .
\label{eq:quartic}
\end{equation}
For $\alpha=0$, Eq.~\eqref{eq:quartic} factorizes as $r \bigl( r^{3}+\ell^{2}r-2\ell^{2}M \bigr)=0$, so the roots are $r=0$ and the positive real solution $r=r_{0}$ of
\begin{equation}
r_{0}^{3}+\ell^{2}r_{0}-2\ell^{2}M=0 .
\label{eq:cubic}
\end{equation}
The trivial root $r=0$ becomes a small but finite horizon when $\alpha \neq 0$, and $r=r_{0}$ is shifted perturbatively. Expanding the root near $r=0$ as $r_{-}=\alpha r_{-}^{(1)}+\mathcal{O}(\alpha^{2})$ and solving order by order gives
\begin{equation}
r_{-} \;=\; \frac{\alpha}{2} + \mathcal{O}(\alpha^{2}) .
\end{equation}
For the outer horizon we expand around $r=r_{0}$, first we expand the root perturbatively about $r=r_0$,
\begin{equation*}
    r_{h}=r_{0}+\alpha\,r_{h}^{(1)}+\mathcal{O}(\alpha^{2}) \ ,
\end{equation*}
and insert this into $F(r,\alpha)=0$ and the Taylor-expand of $F$ to first order in $\alpha$
\begin{equation*}
    0=F(r_{h},\alpha)=F(r_{0},0) +\alpha\Big[r_{h}^{(1)}\partial_{r}F(r_{0},0)+\partial_{\alpha}F(r_{0},0)\Big] +\mathcal{O}(\alpha^{2}) \ .
\end{equation*}
Since $F(r_{0},0)=0$ by definition, the $\mathcal{O}(\alpha)$ equation is
\begin{equation*}
r_{h}^{(1)}\partial_{r}F(r_{0},0)+\partial_{\alpha}F(r_{0},0)=0
\quad\Longrightarrow\quad
r_{h}^{(1)}=-\frac{\partial_{\alpha}F(r_{0},0)}{\partial_{r}F(r_{0},0)} \ .
\end{equation*}
 By computing the derivatives $\partial_{\alpha}F(r,\alpha)=\ell^{2}M$ and $
\partial_{r}F(r,0) = 4r^{3}+2\ell^{2}r-2\ell^{2}M$ and by simplifying the denominator, we get
\begin{equation*}
    r_{h}^{(1)}=-\frac{\ell^{2}M}{3r_{0}^{3}+\ell^{2}r_{0}} \ .
\end{equation*}
Using the constraints from the cubic equation for $r_0$, reveals $2\ell^{2}M = r_{0}^{3}+\ell^{2}r_{0}$, one obtains 
\begin{equation}
r_{h}^{(1)} = -\frac{r_{0}^{2}+\ell^{2}}{2\,(3r_{0}^{2}+\ell^{2})} \, .
\end{equation}
Thus, the outer horizon reads
\begin{equation}
r_{h} = r_{0} - \alpha\,\frac{r_{0}^{2}+\ell^{2}}{2(3r_{0}^{2}+\ell^{2})} + \mathcal{O}(\alpha^{2}) ,
\end{equation}
where $r_{0}$ is the positive solution of Eq.~\eqref{eq:cubic}. This provides a compact perturbative description of both the inner and outer horizons in terms of the physical parameters $M, \ell$, and $\alpha$. One can note that the introduction of $\alpha$ forces the black hole to have two values of the horizon, hence it works like an extra degree of freedom for the black hole. It is apparent from the Fig. \ref{Fig:Metric Plot} that on increasing the value of $\alpha$, black holes move towards extremality and after that extremal limit of $\alpha$ we have naked singularity. Hence for a physical theory, the upper limit of $\alpha$ is set by the mass of the black hole. For our further discussion, we have taken the value of $\alpha$ which is less than the extremal limit.

In Fig.~\ref{Fig:Metric Plot} we display $f(r)$ for $M=0.5$ and $M=1$ and several values of $\alpha$. For $\alpha=0.20$ (red curve) the lapse function admits two zeroes: a small inner root $r_{-}\approx\alpha/2$ and a larger outer horizon $r_{h}$ close to $r_{0}$. As $\alpha$ increases, the inner horizon shifts outward linearly with $\alpha$, while the outer horizon decreases slightly due to the negative correction proportional to $\alpha$. The overall effect is a shrinking of the physical black hole size as the correction grows. A similar pattern holds for larger masses.

\subsection*{Thermodynamics}
The mass of the black hole, interpreted as its enthalpy in the context of extended thermodynamics, can be determined by solving the condition $f(r_h) = 0$, yielding
\begin{equation}\label{Mass_expr}
   M =  \frac{\pi  r_h \left(\pi ^2 \alpha ^2+64 r_h^2\right) \left(3-\Lambda  r_h^2\right)}{12 \left(\pi ^2 \alpha ^2+64 r_h^2\right) \cot ^{-1}\left(\frac{\pi  \alpha }{8 r_h}\right)-96 \pi  \alpha  r_h} \ ,
\end{equation}
where $r_h$ is the outer horizon. The Hawking temperature of the black hole is defined via the surface gravity as
\begin{equation}\label{Temp_expr}
    T = \frac{256 \alpha  r_h^2 \left(\Lambda  r_h^2-3\right)}{3 \left(\pi ^2 \alpha ^2+64 r_h^2\right) \left(\left(\pi ^2 \alpha ^2+64 r_h^2\right) \cot ^{-1}\left(\frac{\pi  \alpha }{8 r_h}\right)-8 \pi  \alpha  r_h\right)}+\frac{1-\Lambda  r_h^2}{4 \pi  r_h} \ .
\end{equation}
In the extended phase space formulation of AdS black holes, the cosmological constant $\Lambda$ is interpreted as a thermodynamic pressure, as $ P = -\Lambda/8\pi$. The first law of thermodynamics is
\begin{equation}\label{First_Law_Lorentz}
    dM = T \, dS +V \, dP + \aleph \, d\alpha \ .
\end{equation}
Here, $\aleph$ is conjugate to the noncommutative parameter $\alpha$. Using Eq.~\eqref{Mass_expr}, Eq.~\eqref{Temp_expr} and Eq.~\eqref{First_Law_Lorentz}, one can compute the expression of entropy as 
\begin{eqnarray}\label{Entropy_expr}
    S = \pi  r_h^2 +\pi  \alpha  r_h + \mathcal{O}(\alpha^2) \ .
\end{eqnarray}
 The entropy is not the Bekenstein-Hawking entropy; rather, there is some correction in that, and for that reason, we intentionally denote it by $S$. \cite{Wang:2024jlj, Tan:2024jkj, Wang:2024jtp} showed that the first law should be modified to get the entropy exactly the Bekenstein-Hawking entropy. We first discuss the thermodynamic quantities that need to be modified so that the first law is written in its conventional form with entropy as Bekenstein–Hawking entropy. Since the stress energy tensor is a function of mass, it is believed that mass should be changed. Our aim is to compute the expression of how the mass should change to get the Bekenstein–Hawking entropy in the first law of thermodynamics. We assume the metric function is only a function of mass and $r$; to compute the mass of the black hole, we have to use the condition $f(M,r_h) = 0 $. The first law is modified as
 \begin{eqnarray}\label{First law with corrected mass}
      d \Tilde{M} = TdS_{\rm BH} + V_c dP +\aleph_c d\alpha \ .
 \end{eqnarray}
 In Eq.~\eqref{First_Law_Lorentz}, the entropy is the corrected entropy, volume is the corrected volume, and in Eq.~\eqref{First law with corrected mass}, the entropy is the Bekenstein–Hawking entropy; volume must be the volume we usually have, i.e., proportional to $r_h^3$ and independent of $\alpha$, but the mass will be corrected. Also, both the masses are related to the deformation parameter $\mathcal{W}$ with the relation
 \begin{equation}\label{Relation of both mass}
     d\Tilde{M} = \mathcal{W}(r_h)dM \ .
 \end{equation}
By differentiating the metric function with respect to the mass parameter and employing the definition of temperature, together with the relations between the corrected and standard mass forms~\eqref{First law with corrected mass} and Eq.~\eqref{Relation of both mass} and finally using the metric ansatz~\eqref{Ansatz}, and using the $t-t$ component of the Einstein tensor, we have the deformation parameter relation as  \begin{eqnarray}\label{W_definition}
    \mathcal{W}(r_h) = 1 + \int_{r_h}^{\infty} 4\pi r^2 \frac{\partial T_t^t}{\partial M}dr \ .
\end{eqnarray}
The analysis supports the assumption that when the energy-momentum tensor of the gravitational system depends explicitly on the black hole mass $M$, the conventional form of the first law of black hole thermodynamics is no longer preserved. This deviation reflects a modification of the underlying thermodynamic structure, necessitating a revised formulation of the first law in such contexts. Finally, using the Lorentzian case as in Eq.~\eqref{RhoL}, the deformation parameter is 
\begin{eqnarray}\label{WofLorentz}
    \mathcal{W}(r_h) = \frac{2}{\pi } \cot ^{-1}\left(\frac{\pi  \alpha }{8 r_h}\right)-\frac{16 \alpha  r_h}{\pi ^2 \alpha ^2+64 r_h^2} \ . 
\end{eqnarray}
 Now, using this, one can easily verify using this function~\eqref{WofLorentz}, that the entropy is the Bekenstein-Hawking entropy. Also, we have checked the volume with the corrected first law takes the form $4/3(\pi r_h^3)$. Interestingly, this procedure showed that we started to get the entropy having the form with no $\alpha$-dependent term, and to our surprise, we also get the thermodynamic volume, which also has no $\alpha$-dependent corrections.

\section{Thermodynamics Universality}

Now, we analyze the thermodynamic behavior and investigate the validity of universal relations for the black hole solutions under consideration. To probe the robustness of the proposed universality relation, we introduce a perturbative deformation to the gravitational action, with the perturbation term taken to be proportional to the cosmological constant. This perturbative extension affects not only the spacetime metric but also leads to modifications in the associated thermodynamic quantities. Specifically, we introduce a small parameter $\varrho$, which controls the strength of the perturbation and allows for a systematic expansion around the unperturbed solution. The perturbation is scaled appropriately with respect to the cosmological constant to ensure consistency with the extended phase space framework. Using this framework, we compute the leading-order corrections to the black hole mass, as well as other key thermodynamic variables such as temperature, and pressure in the terms of entropy\footnote{To represent in the terms of entropy we first invert Eq.~\eqref{Entropy_expr}, but it is not possible to invert for more than $\alpha$ order. Upto the first order of $\alpha$ and inverting Eq.~\eqref{Entropy_expr} the horizon radius in the terms of entropy $S$ is 
\begin{eqnarray}
    r_h=\frac{1}{2} \left(\sqrt{\alpha ^2+\frac{4 S}{\pi }}-\alpha \right) \ .
\end{eqnarray}}. These corrected expressions enable a precise evaluation of the universality relation \cite{Goon:2019faz, Anand:2025btp}. The analytical form of the perturbed quantities facilitates a direct verification of the universality relation in the modified setup. The perturbed mass is 
\begin{eqnarray}
    M (\varrho) = \frac{\left(\sqrt{\pi  \alpha ^2+4 S}-\sqrt{\pi } \alpha \right)^4 \left(\pi  \left(6-\alpha ^2 \Lambda  (\varrho +1)\right)+\sqrt{\pi } \alpha  \Lambda  (\varrho +1) \sqrt{\pi  \alpha ^2+4 S}-2 \Lambda  S (\varrho +1)\right)}{\pi ^3 \left(\pi ^2-144\right) \alpha ^3+144 \pi ^{5/2} \alpha ^2 \sqrt{\pi  \alpha ^2+4 S}+96 \pi ^{3/2} S \sqrt{\pi  \alpha ^2+4 S}-384 \pi ^2 \alpha  S} \nonumber \ .
\end{eqnarray}
Using this, we can easily compute 
\begin{eqnarray}\label{dmdvarrho}
    \frac{\partial M}{\partial \varrho} = -\frac{\Lambda  S^{3/2}}{6 \pi ^{3/2}}+\frac{\alpha  \Lambda  S}{6 \pi }-\frac{7 \alpha ^2 \left(\Lambda  \sqrt{S}\right)}{48 \sqrt{\pi }}+\frac{\left(48+\pi ^2\right) \alpha ^3 \Lambda }{1152}+\mathcal{O}\left(\alpha ^4\right) \ .
\end{eqnarray}
Now, we compute the perturbed temperature as 
\begin{eqnarray}\label{Temp_Perturbed}
    T(\varrho) = \frac{T_{\text{Num.}}}{T_{\text{Den.}}} \ ,
\end{eqnarray}
where 
\begin{eqnarray}\label{TNUMDENM}
    T_{\text{Num.}} &=& \pi ^{9/2} \alpha ^3 \left(\alpha ^2 \Lambda  (\varrho +1)-4\right)+32 \pi ^{5/2} \alpha ^3 \left(6-5 \alpha ^2 \Lambda  (\varrho +1)\right)+96 \Lambda  S^2 (\varrho +1) \sqrt{\pi  \alpha ^2+4 S} \nonumber \\
    && -608 \sqrt{\pi } \alpha  \Lambda  S^2 (\varrho +1)+2 \pi ^{7/2} \alpha ^3 \Lambda  S (\varrho +1)+32 \pi ^2 \alpha ^2 \sqrt{\pi  \alpha ^2+4 S} \left(5 \alpha ^2 \Lambda  (\varrho +1)-6\right) \nonumber \\
    &&+32 \pi  S \sqrt{\pi  \alpha ^2+4 S} \left(13 \alpha ^2 \Lambda  (\varrho +1)-3\right)+32 \pi ^{3/2} \alpha  S \left(15-23 \alpha ^2 \Lambda  (\varrho +1)\right)\nonumber \\
    && -\pi ^4 \alpha ^4 \Lambda  (\varrho +1) \sqrt{\pi  \alpha ^2+4 S} \\
    T_{\text{Den.}}&=& 2 \pi ^{3/2} \left(\sqrt{\pi } \alpha -\sqrt{\pi  \alpha ^2+4 S}\right) \Big(\pi ^{3/2} \left(\pi ^2-144\right) \alpha ^3+144 \pi  \alpha ^2 \sqrt{\pi  \alpha ^2+4 S}+96 S \sqrt{\pi  \alpha ^2+4 S} \nonumber\\
    &&-384 \sqrt{\pi } \alpha  S\Big) \nonumber
\end{eqnarray}
Using the expression of $M(\varrho)$ and Eq.~\eqref{Mass_expr}, the perturbed parameter is 
\begin{eqnarray}\label{Epsilon}
    \varrho &=& \frac{1}{\Lambda  \left(\sqrt{\pi  \alpha ^2+4 S}-\sqrt{\pi } \alpha \right)^6} \Bigg[32 \pi ^3 \alpha ^3 \left(\alpha ^3 (-\Lambda )+3 \alpha +9 M\right)-2 \pi ^5 \alpha ^3 M+192 \pi ^2 \alpha  S \Big(\alpha ^3 (-\Lambda ) \nonumber \\
    && +2 \alpha +4 M\Big)+64 \pi ^{3/2} S \sqrt{\pi  \alpha ^2+4 S} \left(2 \alpha ^3 \Lambda -3 (\alpha +M)\right)+32 \pi ^{5/2} \alpha ^2 \sqrt{\pi  \alpha ^2+4 S} \Big(\alpha ^3 \Lambda  \nonumber \\ 
    && -3 \alpha -9 M\Big)-64 \Lambda  S^3+96 \pi  S^2 \left(2-3 \alpha ^2 \Lambda \right)+96 \sqrt{\pi \alpha  \Lambda  S^2 \sqrt{\pi } \alpha ^2+4 S}\Bigg] \ .
\end{eqnarray}
Again, using Eq.~\eqref{Temp_Perturbed}, Eq.~\eqref{TNUMDENM} and Eq.~\eqref{Epsilon} we have 
\begin{eqnarray}\label{Tepslondvarrho}
    T(\varrho) \frac{\partial S}{\partial \varrho } &=& \frac{\Lambda  \left(-\pi ^{3/2} \alpha ^3+\pi  \alpha ^2 \sqrt{\pi  \alpha ^2+4 S}+2 S \sqrt{\pi  \alpha ^2+4 S}-4 \sqrt{\pi } \alpha  S\right)}{24 \pi ^{3/2}} \nonumber \\
    &=& \frac{\Lambda  S^{3/2}}{6 \pi ^{3/2}}-\frac{\alpha  \Lambda  S}{6 \pi }+\frac{5 \alpha ^2 \Lambda  \sqrt{S}}{48 \sqrt{\pi }}-\frac{\alpha ^3 \Lambda }{24}+ \mathcal{O}\left(\alpha ^4\right) \ .
\end{eqnarray}

Finally, Eq.~\eqref{dmdvarrho} and Eq.~\eqref{Tepslondvarrho}, we get a thermodynamic identity at extremality. Specifically, we compute the variation of the extremal black hole mass $M_{\text{ext}}$ with respect to a perturbative parameter $\varrho$, up to third order of the non-commutative parameter $\alpha$ and then $T(\varrho) \frac{\partial S}{\partial \varrho}$. By comparing these results, one verifies that it satisfies the extremality condition:
\begin{eqnarray}\label{extremality condition}
    \frac{\partial M_{\text{ext}}}{\partial \varrho} = -\lim_{M \to M_{\text{ext}}} T(\varrho) \frac{\partial S}{\partial \varrho} \ ,
\end{eqnarray}
confirming the thermodynamic relation proposed in earlier works \cite{Goon:2019faz}. One notices that the extremality condition only verifies up to order $\varrho$, which is expected because, due to computational constraints, we only consider entropy up to first order and invert them to get the horizon radius in terms of entropy. It is expected that if one considers the higher order $\alpha$, and inverts to get the horizon radius, one expects to satisfy the generalized extremality as emphasized in \cite{Anand:2025btp}; this identity may require generalization in scenarios where the entropy deviates from the quadratic area scaling. In the higher-order scenario, the entropy follows the logarithmic form. In such cases, the simple proportionality between entropy and the horizon area is replaced by a more complex functional dependence. Hence, the perturbative approach not only confirms the extremality condition in conventional settings but also provides a powerful tool to test the thermodynamic structure of black holes across a wide class of gravitational theories.

\section{Thermodynamic stability Analysis}\label{Thermodynamic phase}

In this section, we study the thermodynamic stability of Schwarzschild-AdS black holes in the extended phase space, incorporating quantum fuzziness. The black hole temperature is expressed in terms of the thermodynamic pressure via the cosmological constant, which serves as the equation of state with horizon radius $r_h$ as the key variable. The equation of state is
\begin{eqnarray}\label{Temperature}
  T(r_h, P) = \frac{1}{4 \pi  r_h} +2 P r_h -\frac{256 \alpha  r_h^2 \left(8 \pi  P r_h^2+3\right)}{3 \left(\pi ^2 \alpha ^2+64 r_h^2\right) \left(\left(\pi ^2 \alpha ^2+64 r_h^2\right) \cot ^{-1}\left(\frac{\pi  \alpha }{8 r_h}\right)-8 \pi  \alpha  r_h\right)}
\end{eqnarray}

\begin{figure}[h!]
    \begin{center}
        \includegraphics[scale=.55]{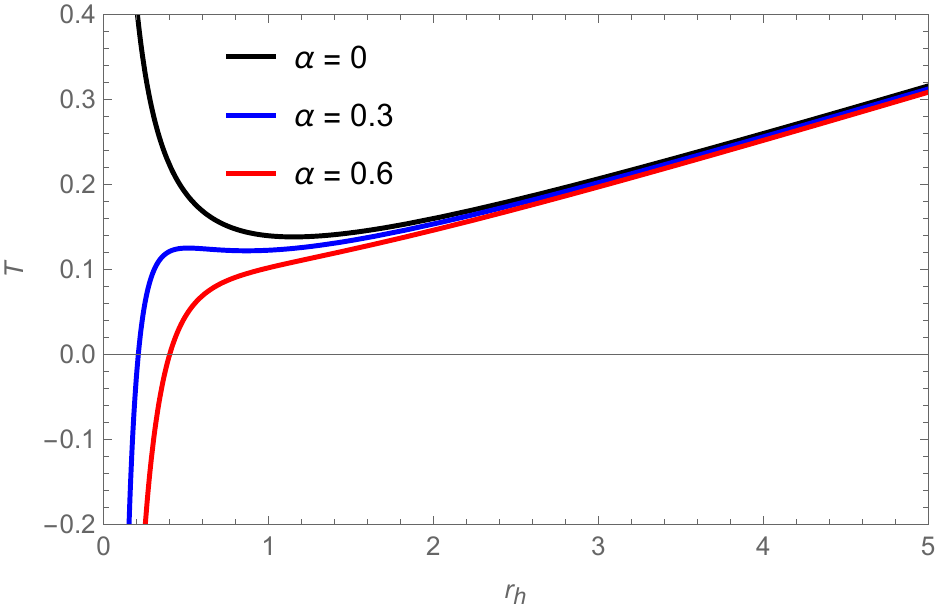} 
    \end{center}
    \caption{The behavior of black hole temperature $T$ with horizon radius $r_{h}$ for fixed values of $\alpha$ and pressure $P=0.03$.}
    \label{T_vs_rh}
\end{figure}

We investigate the effect of the non-commutative parameter $\alpha$ on the thermodynamic stability of non-commutative Schwarzschild AdS black holes. In Fig.~\ref{T_vs_rh}, we plot the Hawking temperature $T$ as a function of the horizon radius $r_h$ at fixed pressure $P$ for different values of $\alpha$. For $\alpha=0$, the temperature exhibits a minimum below which $T$ decreases with increasing $r_h$, indicating a negative specific heat $C_P$. This negative $C_P$ signals that the black hole cannot be in thermodynamic equilibrium with a heat bath, corresponding to thermodynamic instability. Owing to the similarity between the equation of state of RN-AdS black holes and that of a Van der Waals fluid, the authors of \cite{Spallucci:2013osa} proposed that the experimental results for VdW fluids could be extended to black holes. They applied the Maxwell equal area law to RN-AdS black holes and demonstrated that the unphysical region can be replaced by an isothermal segment, interpreted as a mixture of ``liquid'' and ``gas'' black hole phases. A similar analysis was performed by \cite{Wei:2019uqg}, who replaced the unphysical region with a coexistence curve computed via the Maxwell construction and argued for a universal microstructure of black holes, independent of the black hole charge. However, the use of reduced variables, $\tilde T = T/T_c$ and $\tilde V = V/V_c$, implicitly introduces a charge dependence, since the critical quantities scale as $T_c \sim 1/Q$ and $V_c \sim Q^3$.

For the non-commutative Schwarzschild AdS black holes studied here, our equation of state does not resemble that of a VdW fluid, and there is no experimental guidance to justify assuming coexistence phases. Therefore, rather than replacing the unstable region with a coexistence curve, we ask a more fundamental question: Can the black hole be rendered thermodynamically stable by tuning the non-commutative parameter $\alpha$? Although the extremum points of the temperature cannot be obtained analytically from Eq.~\ref{Temperature}, numerical analysis shows that increasing $\alpha$ removes these unstable extrema. As evident from the red curve in Fig.~\ref{T_vs_rh}, the black hole temperature $T$ becomes a monotonically increasing function of horizon $r_h$, corresponding to positive specific heat $C_P > 0$ and thermodynamic stability.

Further, it is straightforward to compute the equation of state using Eq.~\ref{Temperature} as a function of horizon $r_h$ and thermodynamic temperature $T$ as,
\begin{equation}\label{eq_of_state}
P(r_h, T) = \frac{4 \pi  T r_h-1}{8 \pi  r_h^2} + \frac{256 r_h \left(\alpha +2 \pi  \alpha  T r_h\right)}{3 \left(\pi ^2 \alpha ^2+64 r_h^2\right){}^2 \cot ^{-1}\left(\frac{\pi  \alpha }{8 r_h}\right)-8 \left(3 \pi ^3 \alpha ^3 r_h+320 \pi  \alpha  r_h^3\right)}  \ .
\end{equation}

\begin{figure}[h!]
\begin{center}
\includegraphics[scale=.5]{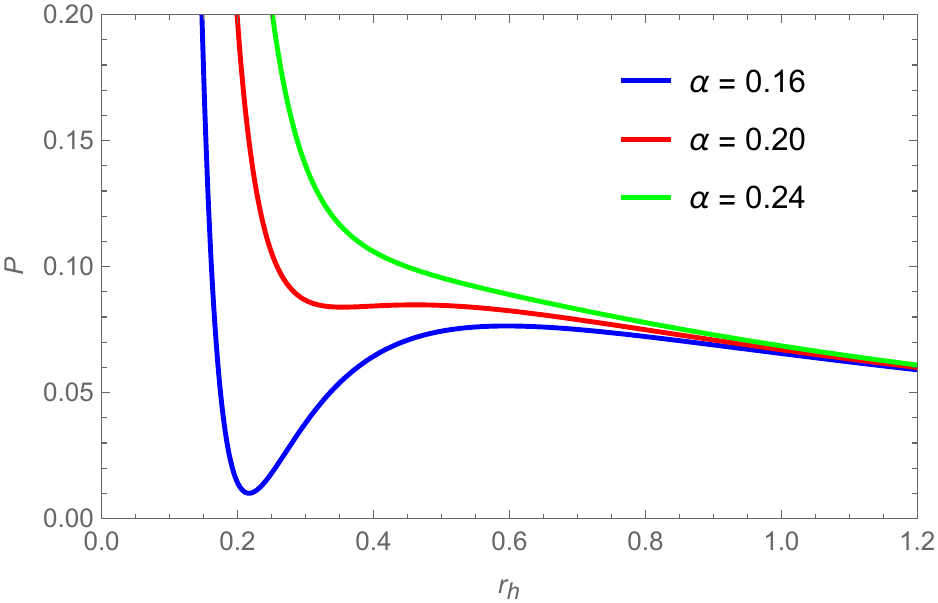}
\end{center}
\caption{The behavior of pressure $P$ with horizon $r_h$ for fixed values of $\alpha$ and temperature $T=0.2$.}
\label{P_vs_rh}
\end{figure}

Again, by plotting Eq.\eqref{eq_of_state}, we can analyze the thermodynamic behavior of pressure $P$ as a function of horizon $r_h$, shown in Fig.\ref{P_vs_rh}. For black holes incorporating quantum fuzziness via non-commutative geometry, the temperature isotherms display rich thermodynamic behavior. It is clear from the fig \ref{P_vs_rh} that the oscillating part which corresponds to negative compressibility is removed by increasing $\alpha$.

Finally, we study the thermodynamic stability of black holes using the specific heat at constant pressure, $C_P$, formalism within the extended phase space. The specific heat at constant pressure is 
\begin{eqnarray}
    C_P &=&\frac{\pi \left( 24 \left(\alpha + 2 r_h\right) \left(\alpha^{2} + 4 r_h^{2}\right) - \pi^{2} \alpha^{3} \right) \mathcal{H}}{96 r_h^{2}\;\partial_{r_h}\mathcal{H}}, 
\end{eqnarray}
where,
\begin{eqnarray}
    \mathcal{H} &=& \frac{1}{4 \pi r_h} + 2 P r_h - \frac{256 \alpha r_h^{2} \left( 8 \pi P r_h^{2} + 3 \right)} {3 \left( \pi^{2} \alpha^{2} + 64 r_h^{2} \right)  \left( \left( \pi^{2} \alpha^{2} + 64 r_h^{2} \right)        \cot^{-1}\!\left( \frac{\pi \alpha}{8 r_h} \right) - 8 \pi \alpha r_h \right)} \ .
\end{eqnarray}

\begin{figure}[h!]
    \begin{center}
        \includegraphics[scale=.5]{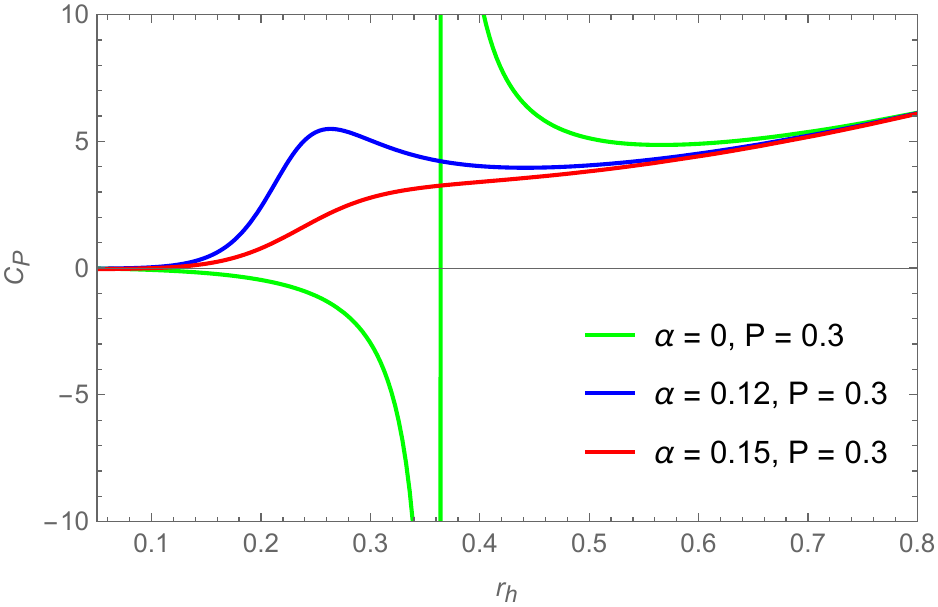} 
    \end{center}
    \caption{Behavior of $C_P$ with horizon $r_h$ for fixed values of $\alpha$ and pressure $P$.}
    \label{Cp vs rh}
\end{figure}

The behavior of the specific heat at constant pressure $C_{P}$ as a function of the horizon radius $r_h$ for the non-commutative Schwarzschild-AdS black hole is shown in Fig.~\ref{Cp vs rh}.
 For small $\alpha$, $C_{P}$ remains negative at small $r_h$, indicating thermal instability in this regime~\cite{Tan:2024jkj}. However, as $\alpha$ increases, the region of negative $C_{P}$ progressively shrinks, and the specific heat becomes predominantly positive. This demonstrates that the non-commutative corrections act to stabilize small black holes by suppressing the unstable branch.

\section{Thermodynamic Scalar Curvature} \label{Sec:Thermodynamic Curvature}

Probing the microscopic structure of black holes in non-commutative geometry has provided profound modifications to classical gravitational frameworks, particularly in the ultraviolet regime where quantum gravitational effects are expected to emerge. We employ the Ruppeiner geometric approach to thermodynamic fluctuations. The entropy, as per Boltzmann's definition, is proportional to $ln$ of the number of accessible microstates ($\omega$) with the proportionality constant as the Boltzmann constant. Without loss of generality, the Boltzmann constant can be put to unity. We consider a thermodynamic system $\mathcal{I}_0$, described by two fluctuating thermodynamic degrees of freedom, let's say $y^1$ and $y^2$, which remains in equilibrium with an embedded subsystem $\mathcal{I}$. The statistical behavior of such a system is encoded in the probability density function $\mathcal{P}(y^1, y^2)$, representing the likelihood of observing a fluctuation within the infinitesimal domain $(y^1, y^2) \to (y^1 + dy^1, y^2 + dy^2)$. According to the statistical interpretation of thermodynamics, this probability is intimately linked to the number of accessible microstates. At equilibrium, the entropy reaches an extremum—specifically, a maximum—consistent with the second law of thermodynamics. Consequently, all spontaneous fluctuations are expected to occur in the vicinity of this maximum. To quantify these deviations, one performs a Taylor expansion of the entropy function about its equilibrium configuration. When truncated at second order, this expansion leads to a Gaussian approximation for the fluctuation distribution and written as
\begin{equation}\label{fluct_prob}
\mathcal{P} \propto \exp\left( -\frac{1}{2} \delta \ell^2 \right) \ , 
\end{equation}
where $\delta \ell^2$ denotes the squared thermodynamic length, given by
\begin{equation}\label{ruppeiner_metric}
\delta \ell^2 = - \frac{\partial^2 \mathcal{S}}{\partial y^\alpha \partial y^\beta} \delta y^\alpha \delta y^\beta \ , 
\end{equation}
this formulation defines the Ruppeiner metric—a Riemannian metric on the manifold of thermodynamic states. The scalar curvature $\mathcal{R}$ derived from this metric carries information about the nature of microscopic interactions. In particular, the sign of $\mathcal{R}$ distinguishes repulsive/fermionic type ($\mathcal{R} > 0$) from attractive/bosonic type ($\mathcal{R} < 0$) interactions, while its divergence signals critical behavior, such as second-order phase transitions. For black holes influenced by non-commutative geometry, the Ruppeiner curvature encodes how the smearing of matter distributions—resulting from the underlying non-locality modifies the statistical interactions between microstates. Studies have shown that non-commutative black holes can exhibit regular horizons, modified Hawking temperatures, and novel thermodynamic behavior not present in their commutative counterparts. The geometric approach via Ruppeiner analysis complements these findings by offering an interpretation about the nature of microstructures for such phenomena. Initially formulated for BTZ black holes~\cite{Cai:1998ep} and extended to various AdS spacetimes~\cite{Shen:2005nu, Wei:2019uqg, mrugala1993, Singh:2020tkf, Singh:2023hit, Singh:2025ueu}, Ruppeiner geometry has proven effective across systems with higher-curvature corrections and non-minimal couplings~\cite{Dehyadegari, Sheykhi, Miao3, Chen}. Recent formulations place it within a contact geometric framework~\cite{hermann, mrugala1996, contactBH}, where thermodynamic processes evolve along Legendre submanifolds. This covariant approach unifies fluctuations, dualities, and geometric structures in black hole thermodynamics modified by non-commutativity. In non-commutative black hole thermodynamics, constructing a meaningful geometric framework can be challenging due to limited independent thermodynamic variables—especially in neutral, non-rotating cases where all quantities depend solely on the horizon radius. However, extended thermodynamics, which treats the cosmological constant as pressure $P$, introduces new degrees of freedom, enabling a consistent formulation of thermodynamic geometry even in such simplified settings. Using enthalpy $H(S, P)$ as the thermodynamic potential allows the construction of Ruppeiner metrics in the $(S, P)$ space~\cite{Ghosh:2023khd}. Yet, when $S$ and $V$ are not independent—as in non-commutative or non-Abelian configurations—internal energy representations become inadequate. The appropriate choice of fluctuation coordinates depends on the ensemble, and while curvature scalars from different ensembles may differ, they remain consistent when derived from ensemble-compatible metrics~\cite{Bravetti:2012dn}. Curvature divergences often indicate phase transitions, but in some parametrizations, Ruppeiner curvature remains finite despite diverging response functions~\cite{Ghosh:2023khd}. This has led to alternative thermodynamic metrics~\cite{Dolan:2015xta, Singh:2020tkf, Singh:2023hit}, which better capture critical behavior. Notably, the normalized scalar curvature~\cite{Wei:2019uqg} remains effective across systems, including non-commutative black holes. A detailed classification of such metrics and their thermodynamic interpretations is available in~\cite{Wei:2019yvs, ANAND2025101994}.

To investigate microscopic interactions and critical phenomena in non-commutative black holes, Ruppeiner geometry can be effectively constructed within the extended thermodynamic phase space. In this framework, non-commutative effects smear the matter distribution, modifying the black hole’s thermodynamic behavior and introducing novel features into the phase behavior. The Ruppeiner metric in the $(T, V)$-plane is given by~\cite{Wei:2019uqg,Ghosh:2023khd,Wei:2019yvs},
\begin{equation}\label{metric}
d \ell_{\mathcal{R}}^2 = \frac{C_V}{T^2}dT^2 + \frac{1}{T}\left(\frac{\partial P}{\partial V}\right)_T dV^2,
\end{equation}
where $C_V$ is the specific heat at constant volume. For static non-commutative black holes, $C_V = 0$, simplifying the metric to a unique diagonal form that remains consistent with equilibrium thermodynamics. The scalar curvature associated with this metric, often normalized via $C_V$, encodes statistical interactions and signals phase transitions, even in smeared geometries. This normalized Ruppeiner curvature $\mathcal{R_N}$ has been shown to successfully capture critical behavior across various systems, including black holes with quantum corrections~\cite{Wei:2019uqg, Singh:2023hit}. Moreover, the $(T, V)$ formulation derived from the Helmholtz free energy is thermodynamically equivalent to the enthalpy-based $(S, P)$ ensemble, and is particularly well-suited for systems where non-commutativity alters the standard thermodynamic relations. Incorporating non-commutative parameters, along with pressure and charge, extends the phase space and reveals richer microphysical structure.

\begin{figure}[h!]
\begin{center}
\includegraphics[scale=.6]{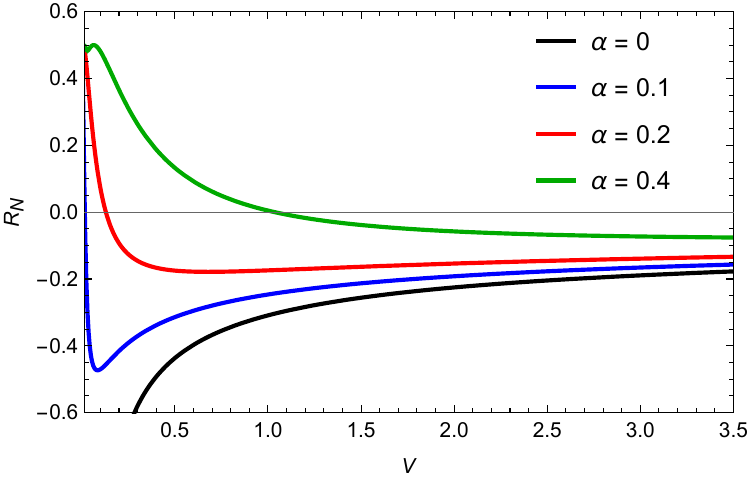}
\end{center}
\caption{The behavior of normalized Ruppeiner curvature $\mathcal{R_N}$ with thermodynamic volume $V$ for fixed black hole temperature $T=1.2$ and $\alpha$ in a non-commutative AdS background.}
\label{RN_vs_V}
\end{figure}

\begin{figure}
    \centering
    \includegraphics[width=0.35\linewidth]{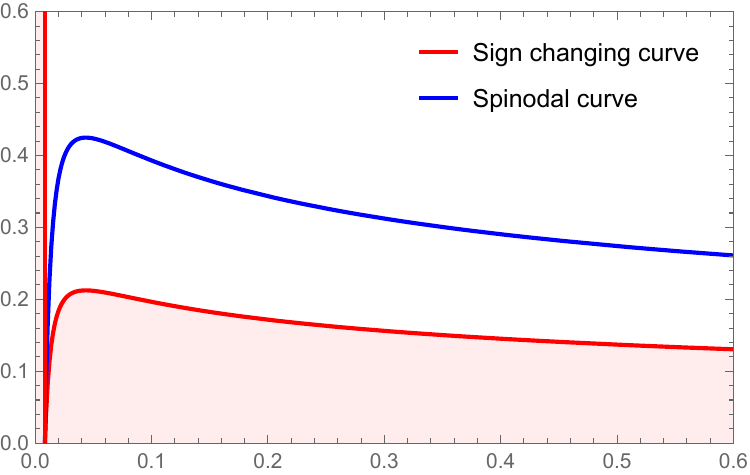}
    \includegraphics[width=0.35\linewidth]{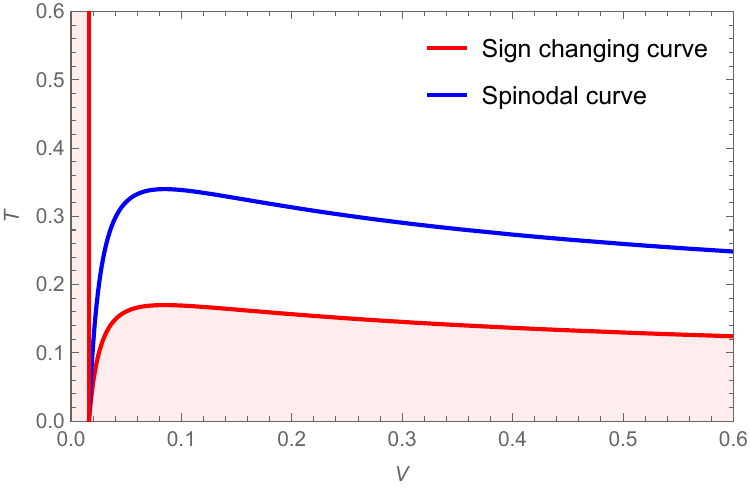}
    \includegraphics[width=0.35\linewidth]{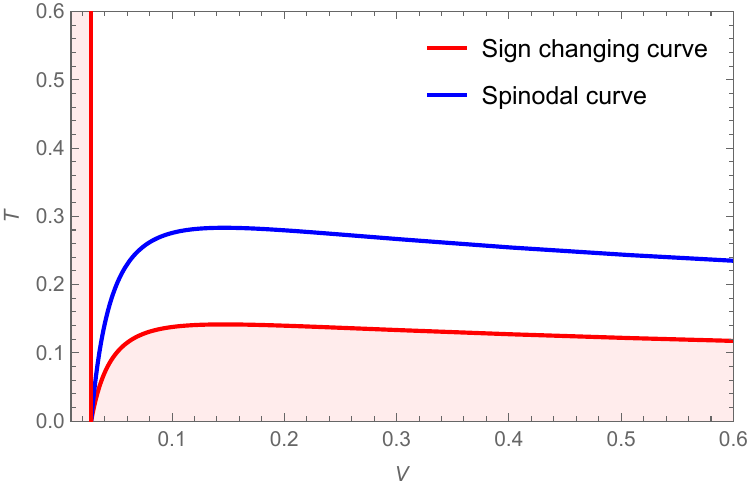}
    \includegraphics[width=0.35\linewidth]{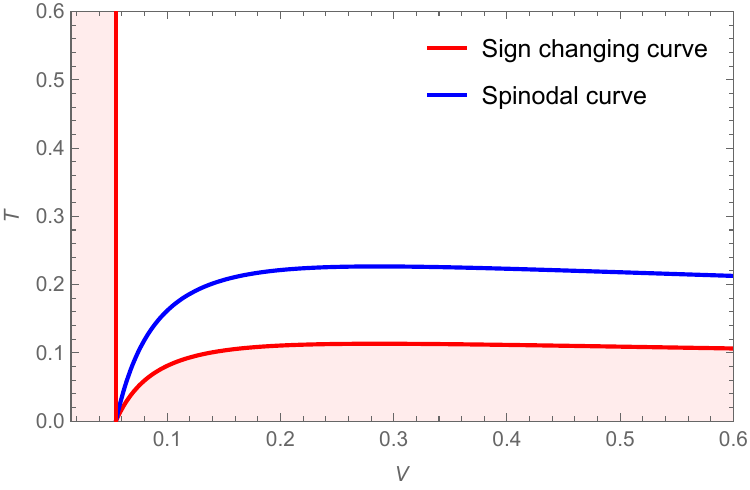}
    \caption{ Characteristic curves for the Schwarzschild AdS black holes in non-commutative geometry for $\alpha =0.08, 0.1, 0.12, 0.15$ respectively. The sign-changing curve is described by the solid red line. The solid blue line represent the spinodal ($T_{\text{div}}$) curve . In the shaded region, the scalar curvature $\mathcal{R_N}>0$; otherwise $\mathcal{R_N}<0$.}
    \label{Sign changing curve}
\end{figure}

To gain further insight into the microscopic structure of the black hole and its interactions, we compute the Ruppeiner curvature scalar using the fluctuation coordinates as $(T, V)$, where $T$ is the temperature and $V$ the thermodynamic volume. The scalar curvature $\mathcal{R_N}$ derived from this geometry serves as an indicator of underlying microscopic interactions. The general analytic expression for the Ruppeiner curvature scalar in non-commutative geometry takes the form,
\begin{eqnarray}
    &&\mathcal{R_N} = \frac{2T \sqrt[3]{6V\pi^2}  -1}{2 \left(T\sqrt[3]{6V\pi^2}  -1\right)^2}-\frac{5   \left(T^2\sqrt[3]{6V\pi^5} \right)}{3^{2/3} \left(T\sqrt[3]{6V\pi^2}-1\right)^3} \alpha+\frac{\pi ^2 T^2 \left(37-12 T\sqrt[3]{6V\pi^2}\right)}{6 \left(T\sqrt[3]{6V\pi^2}-1\right)^4} \alpha ^2\nonumber \\
    && +\frac{\alpha ^3 \left(-18T^4  \sqrt[3]{6V^2\pi^{11}} \left(128+33 \pi ^2\right) +36 \pi ^3 \left(488+33 \pi ^2\right) T^3 \sqrt[3]{V}- \sqrt[3]{36\pi^{7}} \left(4544+99 \pi ^2\right) T^2\right)}{1296 \sqrt[3]{V} \left(\sqrt[3]{6V\pi^{2}}-1\right)^5}+\mathcal{O}(\alpha^4) 
\end{eqnarray}
It is significant to note that in the non-commutative limit $\Theta \to 0$, this curvature reduces to the standard result for Schwarzschild-AdS black holes~\cite{SAdS}. The behavior of $\mathcal{R_N}$ with thermodynamic volume $V$ is shown in Fig.~\ref{RN_vs_V}.
Furthermore, we want to analyze the effect of the perturbation parameter on Ruppeiner curvature and hence on the interactions of microstates of the black hole. Now, the thermodynamic behavior can be characterized by two important curves: the spinodal curve and the sign-changing curve of the normalized Ruppeiner curvature $\mathcal{R_N}$. The spinodal curve is defined by the condition \(\left(\partial P/\partial V \right)_{T} = 0, \)which marks the boundary between locally stable and unstable branches of the equation of state. Along this curve, the normalized Ruppeiner curvature diverges, with the corresponding divergence temperature given by,  
\begin{equation}
T_{\text{div}} = \frac{4 \left(-\sqrt[3]{6} \pi ^6 \alpha ^6+192 \pi ^{10/3} \alpha ^4 V^{2/3}+1152 (6 \pi )^{2/3} \alpha ^2 V^{4/3}-9504 \sqrt[3]{\pi } \alpha  V^{5/3}+2592 \sqrt[3]{6} V^2-198 \sqrt[3]{6} \pi ^3 \alpha ^3 V\right)}{13824 \pi ^{4/3} \alpha ^2 V^{5/3}+10368 (6 \pi )^{2/3} V^{7/3}-20736 \sqrt[3]{6} \pi  \alpha  V^2+6^{2/3} \pi ^{20/3} \alpha ^6 \sqrt[3]{V}-144 \sqrt[3]{6} \pi ^4 \alpha ^4 V}.
\label{Tdiv}
\end{equation}
The sign-changing curve, determined by $\mathcal{R_N} = 0$, separates the $T$–$V$ plane into regions of positive and negative curvature and is expressed as,  
\begin{equation}
T_0 = \frac{2 \left(-\sqrt[3]{6} \pi ^6 \alpha ^6+192 \pi ^{10/3} \alpha ^4 V^{2/3}+1152 (6 \pi )^{2/3} \alpha ^2 V^{4/3}-9504 \sqrt[3]{\pi } \alpha  V^{5/3}+2592 \sqrt[3]{6} V^2-198 \sqrt[3]{6} \pi ^3 \alpha ^3 V\right)}{13824 \pi ^{4/3} \alpha ^2 V^{5/3}+10368 (6 \pi )^{2/3} V^{7/3}-20736 \sqrt[3]{6} \pi  \alpha  V^2+6^{2/3} \pi ^{20/3} \alpha ^6 \sqrt[3]{V}-144 \sqrt[3]{6} \pi ^4 \alpha ^4 V}.
\label{T0}
\end{equation}
It follows that the general relation \(T_{\text{div}} = 2T_0\) continues to hold in the non-commutative case. As shown in Fig.~\ref{Sign changing curve}, the shaded region below and left of the sign-changing curves  corresponds to $\mathcal{R_N} > 0$, while non-shaded region  corresponds to $\mathcal{R_N} < 0$. It is evident from the Fig. \ref{Sign changing curve} that on increasing $\alpha$, the sign changing curve shifts, marking the transition into the attractive micro interactions for earlier repulsive microstructure black holes. 

\section{Conclusion}

In this work, we analyzed the thermodynamics of Schwarzschild-AdS black holes in the framework of non-commutative geometry, where spacetime fuzziness is implemented through Lorentzian smeared matter distributions. Corrected solutions were obtained, and a detailed thermodynamic analysis was carried out. We observe that non-commutativity introduces an extra horizon. On increasing $\alpha$ for a fixed mass, both the horizons coincide, and the black hole becomes extremal. This value of $\alpha$ is the maximum possible value, as further increase in it leads to a naked singularity. Thus, the non-commutative parameter acts like an extra degree of freedom for the black hole. By studying the thermodynamics of the black hole, we observed that the entropy deviates from the Bekenstein–Hawking relation, and the standard first law of black hole thermodynamics is violated. By introducing a correction in the mass term, we established a modified first law consistent with the Bekenstein–Hawking entropy. The study of universality revealed that when the entropy expansion is considered up to order $\alpha$, the universality relation holds only up to that order. Furthermore, as shown in the Appendix~\ref{Appn:Universality Relation For Zero order}, restricting to the zeroth-order expansion implies that the universality condition is satisfied only up to the zeroth power of $\alpha$. 

We observe that the non-commutative parameter not only modifies the microscopic interactions of black holes but also drives them towards thermodynamic stability. For charged AdS black holes \cite{Wei:2019uqg}, stability can be achieved by tuning the black hole charges, since the critical values depend explicitly on the charge configuration. However, such a modification effectively corresponds to considering a different black hole, as distinct charges define distinct solutions. In contrast, the non-commutative parameter $\alpha$ is an external deformation, independent of the black hole charges, and varying $\alpha$ does not change the identity of the black hole itself.

An interesting avenue for future research can be to examine the effects of non-commutative corrections on the thermodynamic topology and photon sphere structure in Schwarzschild-AdS black holes, with a focus on the connections underlying topological charges, phase transitions, and geodesic stability. In strong-gravity conditions, this could provide possible observational signatures and fundamental aspects of black holes thermodynamics.


\section*{Acknowledgements}
A.A. is financially supported by the Institute's postdoctoral fellowship at IITK. A.S. acknowledges support from the Council of Scientific and Industrial Research-Human Resource Development Group (CSIR-HRDG), under the Postdoctoral Research Associate Fellowship, funded through Project No. 03WS(003)/2023-24/EMR-II/ASPIRE.

\appendix

\section{Universality Relation For $\alpha^0$}\label{Appn:Universality Relation For Zero order}

 In this appendix, we show by assuming the entropy expression~\eqref {Entropy_expr} up to $\alpha^0$ and show that the universal relation is verified till $\alpha^0$ order only. So with this assumption, the perturbed mass and temperature are 
 \begin{eqnarray}\label{Alphazero_Mass}
     M(\varrho)&=& \frac{\sqrt{S} \left(\pi ^3 \alpha ^2+64 S\right) (3 \pi -\Lambda  S (\varrho +1))}{12 \sqrt{\pi } \left(\pi ^3 \alpha ^2+64 S\right) \cot ^{-1}\left(\frac{\pi ^{3/2} \alpha }{8 \sqrt{S}}\right)-96 \pi ^2 \alpha  \sqrt{S}} \\ \label{Alphazero_Temp}
     T(\varrho) =&=&  \frac{256 \alpha  S (\Lambda  S (\varrho +1)-3 \pi )}{3 \left(\pi ^3 \alpha ^2+64 S\right) \left(\left(\pi ^3 \alpha ^2+64 S\right) \cot ^{-1}\left(\frac{\pi ^{3/2} \alpha }{8 \sqrt{S}}\right)-8 \pi ^{3/2} \alpha  \sqrt{S}\right)}+\frac{\pi -\Lambda  S (\varrho +1)}{4 \pi ^{3/2} \sqrt{S}} \ .
 \end{eqnarray}
 The perturbation parameter is 
 \begin{eqnarray}\label{Alphazero_Epsilon}
     \varrho =  \frac{\sqrt{S} \left(\frac{96 \pi ^2 \alpha  M}{\pi ^3 \alpha ^2+64 S}-\Lambda  S+3 \pi \right)-12 \sqrt{\pi } M \cot ^{-1}\left(\frac{\pi ^{3/2} \alpha }{8 \sqrt{S}}\right)}{\Lambda  S^{3/2}} \ .
 \end{eqnarray}
Again, using Eq.~\eqref{Alphazero_Temp},  and Eq.~\eqref{Alphazero_Epsilon} we have 
\begin{eqnarray}\label{Tdsdepsilon_alpha0}
    T(\varrho) \frac{\partial S}{\partial \varrho }  
    &=& \frac{\Lambda  S^{3/2}}{6 \pi ^{3/2}} \ .
\end{eqnarray}
Now, we compute the variation of the extremal black hole mass $M_{\text{ext}}$ with respect to a perturbative parameter $\varrho$, we have 
\begin{eqnarray}\label{dmdepsilon_alpha0}
     \frac{\partial M_{\text{ext}}}{\partial \varrho} &=& \left[\frac{96 \pi ^2 \alpha }{64 \Lambda  S^2+\pi ^3 \alpha ^2 \Lambda  S}-\frac{12 \sqrt{\pi } \cot ^{-1}\left(\frac{\pi ^{3/2} \alpha }{8 \sqrt{S}}\right)}{\Lambda  S^{3/2}}\right]^{-1} = -\frac{\Lambda  S^{3/2}}{6 \pi ^{3/2}}-\frac{\alpha  \Lambda  S}{12 \pi }-\frac{\alpha ^2 \Lambda  \sqrt{S}}{24 \sqrt{\pi }} +\mathcal{O}(\alpha^3)  \ .
\end{eqnarray}
Comparing Eq.~\eqref{Tdsdepsilon_alpha0} and Eq.~\eqref{dmdepsilon_alpha0}, it is easy to verify only up to the $\alpha^0$ extremality condition, as in Eq.~\eqref{extremality condition} verified, again this is expected because we only consider entropy up to zeroth order and invert them to get the horizon radius in terms of entropy. 
 

\bibliographystyle{ieeetr}

\bibliography{main.bib}

\end{document}